\documentclass[10pt,twocolumn,twoside]{IEEEtran}
\usepackage{amsmath}
\usepackage{color}
\usepackage{cases}
\usepackage{amsfonts,amsmath,amssymb}
\usepackage{mathrsfs}
\usepackage{cite}
\usepackage{graphicx}
\usepackage{url}
\usepackage{enumerate}
\usepackage{subfigure}
\usepackage{hyperref}
\usepackage{algorithm}
\usepackage{algorithmic}
\usepackage{amsmath,bm}
\usepackage{caption}
\usepackage{float}
\usepackage{stfloats}
\usepackage{multirow}
\usepackage{url}
\usepackage{breakurl}
\usepackage{txfonts}
\usepackage{subfigure}
\usepackage{booktabs}
\usepackage{threeparttable}
\usepackage{multirow}
\usepackage{fancyhdr}

\DeclareMathOperator*{\argmin}{arg\,min}

\newtheorem{remark}{Remark}

\begin{document}
\title{A Manifold Regularized Multi-Task Learning Model for IQ Prediction from Multiple fMRI Paradigms}

\author{\mbox{Li Xiao, Julia M. Stephen, Tony W. Wilson, Vince D. Calhoun, and Yu-Ping Wang}
\thanks{Manuscript received 2019. This work was supported in part by NIH under Grants R01GM109068, R01MH104680, R01MH107354, R01AR059781, R01EB006841, R01EB005846, R01MH103220, R01MH116782, P20GM103472, and in part by NSF under Grant 1539067.}
\thanks{L. Xiao and Y.-P. Wang are with the Department of Biomedical Engineering,
Tulane University, New Orleans, LA 70118, (e-mail: wyp@tulane.edu).}
\thanks{J. M. Stephen and V. D. Calhoun are with the Mind Research Network, Albuquerque, NM 87106. }
\thanks{T. W. Wilson is with the Department of Neurological Sciences, University of Nebraska Medical Center, Omaha, NE 68198.}
\thanks{V. D. Calhoun is with the Department of Electrical and Computer Engineering, University of New Mexico, Albuquerque, NM 87131.}
}
\maketitle

\thispagestyle{fancy}
\fancyhead{}
\lhead{}
\lfoot{\footnotesize{This work has been submitted to the IEEE for possible publication. Copyright may be transferred without notice, after which this version may no longer be accessible.}}
\cfoot{}
\rfoot{}

\begin{abstract}
Multi-modal brain functional connectivity (FC) data have shown great potential for providing insights into individual variations in behavioral and cognitive traits. The joint learning of multi-modal imaging data can utilize the intrinsic association, and thus can boost the learning performance. Although several multi-task based learning models have already been proposed by viewing the feature learning on each modality as one task, most of them ignore the geometric structure information inherent in the modalities, which may play an important role in extracting discriminative features. In this paper, we propose a new manifold regularized multi-task learning model by simultaneously considering between-subject and between-modality relationships. Besides employing a group-sparsity regularizer to jointly select a few common features across multiple tasks (modalities), we design a novel manifold regularizer to preserve the structure information both within and between modalities in our model. This will make our model more adaptive for realistic data analysis.
Our model is then validated on the Philadelphia Neurodevelopmental Cohort dataset, where we regard our modalities as functional MRI (fMRI) data collected under two paradigms. Specifically, we conduct experimental studies on fMRI based FC network data in two task conditions for intelligence quotient (IQ) prediction. The results demonstrate that our proposed model can not only achieve improved prediction performance, but also yield a set of IQ-relevant biomarkers.

\end{abstract}


\section{Introduction}

In recent decades, the human brain functional connectome has emerged as an important ``fingerprint'' to provide insights into individual variations in behavioral and cognitive traits \cite{Sporns1,Cao1,Zuo1}. The functional connectome is quantitatively characterized by a functional connectivity network (FCN) based on graph theory, where the spatially distributed but functionally linked regions-of-interest (ROIs) in the brain represent the nodes and the functional connectivities (FCs) defined as the correlations between the time courses of ROIs represent the edges. The Pearson correlation is widely adopted to measure the FC for its efficiency. It is also worth noting that among neuroimaging studies, functional magnetic resonance imaging (fMRI) is one of the most popular modalities to analyze brain FCNs due to its non-invasiveness, high spatial resolution, and good temporal resolution \cite{Biao1,Calhoun1,Qingbao1}.

Functional connectome-based analyses using fMRI have offered great potential for
understanding the brain-behavior and -cognition relationship, while accounting for variables such as age, gender, intelligence, and disease \cite{Meier2012,logecu2,Lixia2011,Pezoulas2017,Finn1,Song2008,Barch2013,Calhoun2012,Beaty2017,Greene2018,Gao2018}. For instance, Meier \textit{et al.} \cite{Meier2012} constructed FCNs from resting-state fMRI, and then based on these resting-state FCNs, healthy younger and older adults were discriminated by a support vector machine (SVM) classifier. Tian \textit{et al.} \cite{Lixia2011} investigated gender-related differences in the topological organization of resting-state FCNs within the hemispheres on the basis of typical statistical tests. While FCNs are usually constructed from resting-state fMRI, task fMRI based FCs can better explore how individual traits are influenced by brain activity changes induced by trait-related tasks \cite{Buckner2013,Finn2017}.
Calhoun \textit{et al.} \cite{Calhoun2012} used independent component analysis to study fMRI based FCNs from a large group of schizophrenia patients, individuals with bipolar disorder, and healthy controls while performing an auditory oddball task, followed by a multivariate statistical testing framework to infer group differences in properties of identified FCNs. Greene and Gao \textit{et al.} \cite{Greene2018,Gao2018} showed that predictive models built from task fMRI based FC data (e.g., working memory or emotion) can lead to better predictions of fluid intelligence than models built from resting-state fMRI based FC data by the experiments on two large, independent datasets. As such, certain tasks may bring about meaningful findings across subjects with different traits, essentially facilitating biomarker identification beyond what can be found in the resting state.

The majority of previous work has focused on one imaging modality (e.g., resting-state or task fMRI). In neuroimaging research studies, it is common to acquire multi-modal imaging from the same experimental subjects to provide complementary information.
It has also been suggested in \cite{Kaufmann2017,VD-JSui2016} that there is a commonality between different modalities (in brain imaging modality can refer to different functional tasks or different imaging modalities) implicated by the same underlying pathology. To this end,
it is highly desirable to develop an approach for a joint analysis of multiple modalities to boost learning performance. Recently, there have been notable efforts to incorporate multiple modalities in a unified framework for schizophrenia and Alzheimer's disease (AD) diagnosis \cite{newadd1,newadd2,newadd3,Zhang1,Jie1,Lei2018,Zhu1}. Specifically, Zhang \textit{et al.} \cite{Zhang1} proposed a multi-modal multi-task learning model, where multi-task feature learning jointly selected a small number of common features from multiple modalities, and then a multi-modal SVM fused these selected features for both classification and regression. Jie \textit{et al.} \cite{Jie1} and Lei \textit{et al.} \cite{Lei2018} studied a manifold regularized multi-task learning model by viewing the feature learning on each modality as one task. In addition to the group-sparsity regularizer that ensures a few common features to be jointly selected across multiple modalities (tasks), it included the manifold regularizer that preserves the structure information of the data (or called the subject-subject relation) within each single modality. Zhu \textit{et al.} \cite{Zhu1} extended the model in \cite{Jie1} by imposing another two manifold regularizers that preserve the feature-feature relation and response-response relation, respectively. However, all these multi-task based models ignore the subject-subject relation between modalities, which could otherwise improve the final performance.

In this paper, motivated by the work in \cite{Jie1}, we propose a new manifold regularized multi-task learning model, which considers not only the relation of subjects within each single modality but also the relation of subjects between modalities. We extend the model in \cite{Jie1} by replacing the manifold regularizer with a novel one which defines the similarity (or the relation) of subjects by using the Gaussian radial basis function, and particularly the similarity of subjects between different modalities is calculated by propagating the similarity information of subjects within individual modality based on a weighted graph diffusion process. Motivation for this idea is derived from multi-view spectral clustering studied in \cite{Sa1,Lindenbaum1}, and we will introduce it in detail in the next section. From the machine learning point of view, this well-designed manifold regularizer can extract more discriminative features and thereby improving the performance of subsequent prediction. To validate the efficiency and effectiveness of our proposed model, we perform extensive experiments on the publicly available Philadelphia Neurodevelopmental Cohort (PNC) dataset \cite{pncdata1,pncdata2} %a simulated dataset as well as.
Here we predict the continuous-value intelligence quotient (IQ) scores of subjects by using fMRI data in two task conditions (working memory and emotion), and our goal of this study is to investigate which common FCs from the two functional imaging modalities (here our modalities refer to fMRI data collected under two paradigms) contribute most to individual variations in IQ. To be specific, we first construct two FCNs from the two corresponding task fMRI datasets for each subject, respectively. We then regard these FCs as features extracted from the fMRI data and input them into our proposed model for subsequent analysis. It is shown that our proposed model yields improved performance in comparison to the competing models under the metrics of the root mean square error and the correlation coefficient.

The main contributions of this paper are twofold. First, we propose a new manifold regularized multi-task learning model that has two apparent advantages: 1) incorporate complementary information from multiple modalities by jointly learning a small number of common features; and 2) employ a novel manifold regularizer to preserve the structure information of the data both within and between modalities. Second, we apply the proposed model on the real PNC dataset to identify relevant FC biomarkers for IQ prediction using two sets of task fMRI data, and the experimental results show that the proposed model can not only outperform the existing state-of-the-art models, but also discover IQ-relevant predictors that are in accordance with prior studies.

%although changes in brain state may modulate connectivity patterns to
%some degree, an individual¡¯s underlying intrinsic functional architecture
%is reliable enough across sessions and distinct enough from that
%of other individuals to identify him or her from the group regardless
%of how the brain is engaged during imaging.

The remainder of this paper is organized as follows. Section \ref{sec2} describes the existing multi-task based learning models and our proposed new model, respectively. Section \ref{sec3} presents the experimental results on the PNC
data and some discussions. Finally, we conclude this paper in Section \ref{sec4}.

\textit{Notations:} Throughout this paper, uppercase boldface, lowercase boldface, and normal italic letters are used to denote matrices, vectors, and scalars, respectively. The superscript $T$ denotes the transpose of a vector or a matrix. For a matrix $\bm{A}$, we denote its $i$-th row, $j$-th column, $(i,j)$-th entry, and trace as $\bm{A}^{i}$, $\bm{A}_{j}$, $A_{i,j}$, and $\text{tr}(\bm{A})$, respectively. For a vector $\bm{a}$, its $i$-th entry is denoted as
$a(i)$. We further denote the Frobenius norm and $l_{2,1}$-norm of a matrix $\bm{A}$ as $\lVert\bm{A}\rVert_{F}=\sqrt{\sum_{i,j}A_{i,j}^2}$ and $\lVert\bm{A}\rVert_{2,1}=\sum_i\lVert\bm{A}^{i}\rVert_2=\sum_{i}\sqrt{\sum_j A_{i,j}^2}$, respectively.
Let $\mathbb{R}$ denote the set of real numbers.

\section{Methods}\label{sec2}
Multi-task learning (MTL) aims to improve the performance of multiple tasks by exploiting their relationships, particularly when these tasks have some relatedness or commonality \cite{Caruana,Argyriou1}. In \cite{Jie1}, a manifold regularized multi-task learning model has been recently proposed for jointly selecting a small number of common features from multiple modalities and achieved superior performance in AD classification, where each modality was viewed as one task. Importantly, this model considered the structure information of the data within each single modality by adding a manifold regularizer, compared with the classical multi-task learning model. Motivated by the approach in \cite{Jie1}, in this paper we propose a new manifold regularized multi-task learning model, which includes our newly designed manifold regularizer that considers the structure information of the data both within each single modality and between modalities. In this section, we first briefly introduce the existing multi-task based learning models, and subsequently present our proposed model as well as the optimization algorithm.

\subsection{Classical multi-task learning (MTL)}
Assume that there are $M$ different modalities (i.e., tasks). We denote the $m$-th modality as $\bm{X}^{(m)}=[\bm{x}^{(m)}_1,\bm{x}^{(m)}_2,\cdots,\bm{x}^{(m)}_N]^{T}\in\mathbb{R}^{N\times d}$ for $m=1,2,\cdots,M$, where $\bm{x}^{(m)}_i\in\mathbb{R}^{d}$ represents the feature vector of the $i$-th subject in the $m$-th modality, and $d$ and $N$ respectively stand for the numbers of features and subjects. Let $\bm{y}\in\mathbb{R}^{N}$ be the response vector from these subjects, and $\bm{w}^{(m)}\in\mathbb{R}^{d}$ be the regression coefficient vector for the $m$-th modality. Then, the MTL model is to solve the following optimization problem:
\begin{equation}\label{M2R model}
\min_{\bm{W}}\frac{1}{2}\sum_{m=1}^{M}\lVert\bm{y}-\bm{X}^{(m)}\bm{w}^{(m)}\rVert^{2}_{2}+\beta\lVert\bm{W}\rVert_{2,1},
\end{equation}
where $\bm{W}=[\bm{w}^{(1)},\bm{w}^{(2)},\cdots,\bm{w}^{(M)}]\in\mathbb{R}^{d\times M}$ denotes the regression coefficient matrix and $\beta$ is a regularization parameter that balances the tradeoff between residual error and sparsity. The $l_{2,1}$-norm encourages these multiple predictors from different modalities to share similar parameter sparsity patterns, through which the MTL model can result in improved performance for the modality-specific models over training the models separately. It is readily seen that (\ref{M2R model}) is reduced to the least absolute shrinkage and selection operator (LASSO) problem \cite{Tibshirani2011} when the number of modalities equals one.

\subsection{Manifold regularized multi-task learning (M2TL)}
In the classical MTL model above, only the relation between data and the response values is considered, while ignoring the structure information of data, which most likely leads to large deviations. With the expectation that similar subjects should have similar response values, a manifold regularizer that takes into account the subject-subject relation within each single modality is therefore introduced as follows:
\begin{equation}\label{manifold term}
\frac{1}{2}\sum_{i,j}^{N}S^{(m)}_{i,j}\left(\hat{y}^{(m)}(i)-\hat{y}^{(m)}(j)\right)^{2},
\end{equation}
where $\hat{\bm{y}}^{(m)}=\bm{X}^{(m)}\bm{w}^{(m)}=[\hat{y}^{(m)}(1),\hat{y}^{(m)}(2),\cdots,\hat{y}^{(m)}(N)]^{T}\in\mathbb{R}^{N}$ is the estimated response vector and
$\bm{S}^{(m)}=[S^{(m)}_{i,j}]\in\mathbb{R}^{N\times N}$ is the similarity matrix that defines the similarity for each pair of subjects in the $m$-th modality. As for the similarity matrix $\bm{S}^{(m)}$, we construct an adjacency graph by regarding each subject as a node and using the $K$-nearest neighbor rule along with the Gaussian radial basis function to calculate the edge weights as the similarities. If $\bm{x}^{(m)}_i$ is among $K$ nearest neighbors of $\bm{x}^{(m)}_j$ or $\bm{x}^{(m)}_j$ is among $K$ nearest neighbors of $\bm{x}^{(m)}_i$, their similarity $S^{(m)}_{i,j}$ is defined as
\begin{equation}\label{similarity matrix}
S^{(m)}_{i,j}=\exp\left(-\frac{\lVert\bm{x}^{(m)}_i-\bm{x}^{(m)}_j\rVert^{2}_2}{\sigma^{(m)}}\right),
\end{equation}
where $\sigma^{(m)}$ is a free parameter to be fixed empirically as the mean of $\left\{\lVert\bm{x}^{(m)}_i-\bm{x}^{(m)}_j\rVert^{2}_2\right\}_{i\neq j}$; otherwise, $S^{(m)}_{i,j}$ is set to zero, i.e., $S^{(m)}_{i,j}=0$. Let $\bm{L}^{(m)}=\bm{D}^{(m)}-\bm{S}^{(m)}$ be the Laplacian matrix of the graph, where $\bm{D}^{(m)}$ is a diagonal matrix with the diagonal elements being $D^{(m)}_{i,i}=\sum_{j=1}^{N}S^{(m)}_{i,j}$ for $1\leq i\leq N$. Then, (\ref{manifold term}) can be simplified as
\begin{eqnarray}\label{xxxman}
\begin{split}
&\frac{1}{2}\sum_{i,j}^{N}S^{(m)}_{i,j}\left(\hat{y}^{(m)}(i)-\hat{y}^{(m)}(j)\right)^{2}\\
=&(\hat{\bm{y}}^{(m)})^{T}\bm{L}^{(m)}\hat{\bm{y}}^{(m)}
=(\bm{X}^{(m)}\bm{w}^{(m)})^{T}\bm{L}^{(m)}(\bm{X}^{(m)}\bm{w}^{(m)}).
\end{split}
\end{eqnarray}

Based on (\ref{xxxman}), the M2TL model was developed and successfully applied
to AD classification in \cite{Jie1,Zhu1}:
\begin{multline}\label{M3R model}
\min_{\bm{W}}\frac{1}{2}\sum_{m=1}^{M}\lVert\bm{y}-\bm{X}^{(m)}\bm{w}^{(m)}\rVert^{2}_{2}+\beta\lVert\bm{W}\rVert_{2,1}\\
+\gamma\sum_{m=1}^{M}(\bm{X}^{(m)}\bm{w}^{(m)})^{T}\bm{L}^{(m)}(\bm{X}^{(m)}\bm{w}^{(m)}),
\end{multline}
where $\beta$ and $\gamma$ are two regularization parameters.

\subsection{Proposed new M2TL (NM2TL)}
Compared with the MTL model, one appealing property of the M2TL model is that the introduced manifold regularizer $\gamma\sum_{m=1}^{M}(\bm{X}^{(m)}\bm{w}^{(m)})^{T}\bm{L}^{(m)}(\bm{X}^{(m)}\bm{w}^{(m)})$ in (\ref{M3R model}) can preserve the structure information of data. However, it only considers the relation of subjects within each single modality separately, but the important mutual relation of subjects between modalities is ignored. Motivated by this, in this subsection we propose a new M2TL (NM2TL) model that effectively considers both the relation of subjects within the same modality and that between modalities.

\begin{figure}[!h]
  \centering
\includegraphics[width=0.96\columnwidth]{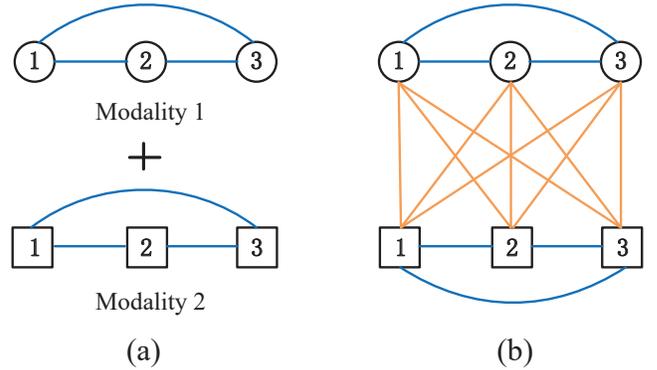}\\
  \caption{The illustration of the relation of data among modalities when $M=2$ and $N=3$. Circles and rectangles respectively represent the subjects in two modalities. Blue connections denote the relation of subjects within each single modality, and orange connections denote the relation of subjects between modalities. (a) and (b) characterize the manifold regularizers in the M2TL model and our proposed NM2TL model, respectively. The M2TL model overlooks the inter-modal connections.}\label{fig_network}
\end{figure}

We first design the following novel manifold regularizer
\begin{equation}\label{newterm}
\mathcal{R}(\bm{W},\gamma,\lambda)=\frac{1}{2}\sum_{p,q}^{M}\sum_{i,j}^{N}\theta_{p,q}S^{(p,q)}_{i,j}\left(\hat{y}^{(p)}(i)-\hat{y}^{(q)}(j)\right)^{2},
\end{equation}
where $\theta_{p,q}$ is a constant such that $\theta_{p,q}=\gamma$ when $p=q$, and $\theta_{p,q}=\lambda$ when $p\neq q$.
Similarly, $\bm{S}^{(p,q)}=[S^{(p,q)}_{i,j}]\in\mathbb{R}^{N\times N}$ is the similarity matrix for each pair of subjects between the $p$-th and $q$-th modalities, i.e., $S^{(p,q)}_{i,j}$ denotes the similarity of the $i$-th subject in the $p$-th modality and the $j$-th subject in the $q$-th modality. Note that $\mathcal{R}(\bm{W},\gamma,\lambda)$ in (\ref{newterm}) is composed of two parts: the first part  $\gamma\frac{1}{2}\sum_{p}^{M}\sum_{i,j}^{N}S^{(p,p)}_{i,j}\left(\hat{y}^{(p)}(i)-\hat{y}^{(p)}(j)\right)^{2}$ preserves the relation of subjects within each single modality; and the second part
$\lambda\frac{1}{2}\sum_{p\neq q}^{M}\sum_{i,j}^{N}S^{(p,q)}_{i,j}\left(\hat{y}^{(p)}(i)-\hat{y}^{(q)}(j)\right)^{2}$ preserves the relation of subjects between modalities. The two free parameters $\gamma$ and $\lambda$ respectively control the effects of the two corresponding parts. In Fig. \ref{fig_network}, the difference between the manifold regularizers in the M2TL and NM2TL models can readily be recognized.

A natural question is how to define the similarity of subjects between two modalities (or nodes from two graphs). We expect that if $\bm{x}^{(q)}_{i}$ and $\bm{x}^{(q)}_{j}$ (i.e., two subjects in the same modality) are similar, the co-occurring subject $\bm{x}^{(p)}_{i}$ corresponding to $\bm{x}^{(q)}_{i}$ should also be similar with $\bm{x}^{(q)}_{j}$. As presented in \cite{Sa1,Lindenbaum1}, the similarity of $\bm{x}^{(p)}_{i}$ and $\bm{x}^{(q)}_{j}$ was calculated in a smooth way by summing over all $N$ co-occurrences, $\bm{x}^{(p)}_{k}$ and $\bm{x}^{(q)}_{k}$ for $1\leq k\leq N$, i.e.,
\begin{equation}
S^{(p,q)}_{i,j}=\sum_{k=1}^{N}S^{(p)}_{i,k}S^{(q)}_{k,j},
\end{equation}
or in matrix form
\begin{equation}
\bm{S}^{(p,q)}=\bm{S}^{(p)}\bm{S}^{(q)},
\end{equation}
where $\bm{S}^{(p)}$ and $\bm{S}^{(q)}$ are the similarity matrices for the $p$-th and $q$-th modalities and calculated by (\ref{similarity matrix}), respectively.
We then put them in a large $MN\times MN$ matrix of the following block-wise form:
\begin{equation}
\bm{S}=\left[
           \begin{array}{cccc}
            \gamma\bm{S}^{(1,1)} & \lambda\bm{S}^{(1,2)} &\cdots &\lambda\bm{S}^{(1,M)} \\
             \lambda\bm{S}^{(2,1)} & \gamma\bm{S}^{(2,2)} &\cdots &\lambda\bm{S}^{(2,M)} \\
             \vdots & \vdots &\cdots &\vdots \\
             \lambda\bm{S}^{(M,1)} & \lambda\bm{S}^{(M,2)} &\cdots &\gamma\bm{S}^{(M,M)}\\
           \end{array}
         \right],
\end{equation}
such that along the diagonal, $\gamma$ is used to tune the within-modality similarity, and off the diagonal, $\lambda$ is used to tune the between-modality similarity. It is obvious that $\bm{S}$ is still symmetrical. Accordingly, by calculating the diagonal matrix $\bm{D}$ where its diagonal elements are $D_{i,i}=\sum_{j}S_{i,j}$ for $1\leq i\leq MN$, we get
\begin{equation}\label{xxlap}
\bm{L}=\bm{D}-\bm{S}.
\end{equation}
Therefore, it is not hard to verify that (\ref{newterm}) can be equivalently expressed as
\begin{eqnarray}\label{newterm2}
\mathcal{R}(\bm{W},\gamma,\lambda)&=&
\left[
    \begin{array}{c}
      \hat{\bm{y}}^{(1)} \\
      \hat{\bm{y}}^{(2)} \\
      \vdots \\
      \hat{\bm{y}}^{(M)} \\
    \end{array}
  \right]^{T}\bm{L}\left[
    \begin{array}{c}
      \hat{\bm{y}}^{(1)} \\
      \hat{\bm{y}}^{(2)} \\
      \vdots \\
      \hat{\bm{y}}^{(M)} \\
    \end{array}
  \right]\nonumber\\
&=&\left[
    \begin{array}{c}
      \bm{X}^{(1)}\bm{w}^{(1)} \\
      \bm{X}^{(2)}\bm{w}^{(2)} \\
      \vdots \\
      \bm{X}^{(M)}\bm{w}^{(M)} \\
    \end{array}
  \right]^{T}\bm{L}\left[
    \begin{array}{c}
      \bm{X}^{(1)}\bm{w}^{(1)} \\
      \bm{X}^{(2)}\bm{w}^{(2)} \\
      \vdots \\
      \bm{X}^{(M)}\bm{w}^{(M)} \\
    \end{array}
  \right].
\end{eqnarray}

Based on the new manifold regularizer in (\ref{newterm2}), the NM2TL model is proposed as follows:
\begin{equation}\label{NM3R model}
\min_{\bm{W}}\frac{1}{2}\sum_{m=1}^{M}\lVert\bm{y}-\bm{X}^{(m)}\bm{w}^{(m)}\rVert^{2}_{2}+\beta\lVert\bm{W}\rVert_{2,1}+\mathcal{R}(\bm{W},\gamma,\lambda),
\end{equation}
where $\beta$, $\gamma$, and $\lambda$ denote control parameters of the respective regularizers. In our NM2TL model (\ref{NM3R model}), the $l_{2,1}$-norm regularizer ensures a sparse set of common features to be jointly learned from multiple modalities, and the manifold regularizer attempts to preserve the structure information of the data both within each single modality and between modalities. Thus, it may extract more discriminative features.

\begin{remark}
More recently, a similar model has been developed in \cite{Meiling1} for identifying the associations between genetic risk factors and multiple neuroimaging modalities under the guidance of the a priori diagnosis information (i.e., AD status). Specifically, a diagnosis-aligned regularizer was introduced to fully explore the relation of subjects with the class level diagnosis information in multi-modal imaging such that subjects from the same class will be close to each other after being mapped into the label space, i.e.,
\begin{equation}\label{teee}
\mathcal{R}(\bm{W})=\sum_{p\leq q}^{M}\sum_{i,j}^{N}S^{(p,q)}_{i,j}\left(\hat{y}^{(p)}(i)-\hat{y}^{(q)}(j)\right)^{2},
\end{equation}
where the similarity $S^{(p,q)}_{i,j}$ is defined as
\begin{equation}\label{sensimilar}
S^{(p,q)}_{i,j}=
  \begin{cases}
    1,       &\!\!\text{if the $i$-th and $j$-th subjects from the same class}\\
    0,  &\!\!\text{otherwise}.
  \end{cases}
\end{equation}
In this way, we can identify a set of common features that are associated with both risk genetic factors and disease status in order to have a better understanding of the biological pathway specific to AD. Our manifold regularizer in the NM2TL model can be clearly distinguished from the above diagnosis-aligned regularizer in a number of aspects: $1)$ Our proposed manifold regularizer is to preserve the geometric structure across modalities such that if the distance of subjects is small, their mapped response values in the label space will also be close. However, the manifold regularizer in (\ref{teee}) aims to preserve the class level diagnosis information. $2)$ In our proposed manifold regularizer, we calculate the similarity of subjects using the Gaussian radial basis function, and particularly the similarity of subjects between different modalities is obtained by propagating the similarity information of subjects within individual modality based on a weighted graph diffusion process.
This similarity measure has been proven to be effective to preserve the structure information of the original data. $3)$ We use two different parameters $\gamma$ and $\lambda$ in our proposed manifold regularizer to balance the relative contribution of the structure information of the data within a single modality and that between modalities, which can result in a better fit to realistic data analysis.
\end{remark}

\subsection{Optimization algorithm}
Clearly, the objective function in (\ref{NM3R model}) is convex but non-differentiable with respect to $\bm{W}$. We write it as a summation of two functions:
\begin{align}
f(\bm{W})&=\frac{1}{2}\sum_{m=1}^{M}\lVert\bm{y}-\bm{X}^{(m)}\bm{w}^{(m)}\rVert^{2}_{2}+\mathcal{R}(\bm{W},\gamma,\lambda),\\ g(\bm{W})&=\beta\lVert\bm{W}\rVert_{2,1},
\end{align}
where $f(\bm{W})$ is convex and differentiable, while $g(\bm{W})$ is convex but non-differentiable.
In this scenario, we optimize $\bm{W}$ in (\ref{NM3R model}) by the commonly used accelerated proximal gradient method \cite{Meiling1,Nesterov1,Parikh1,Zhu2}.

We iteratively update $\bm{W}$ with the following procedure:
\begin{equation}\label{update1}
\bm{W}(t+1)=\argmin_{\bm{W}}\;\Omega_l(\bm{W},\bm{W}(t)),
\end{equation}
where
\begin{multline}
\Omega_l(\bm{W},\bm{W}(t))=f(\bm{W}(t))+\langle\bm{W}-\bm{W}(t), \nabla f(\bm{W}(t))\rangle_F\\
+\frac{1}{2l}\lVert\bm{W}-\bm{W}(t)\rVert^{2}_{F}+g(\bm{W}),
\end{multline}
$\bm{W}(t)$ stands for the value of $\bm{W}$ obtained at the $t$-th iteration, $\langle\bm{W}-\bm{W}(t),\nabla f(\bm{W}(t))\rangle_F=\text{tr}((\bm{W}-\bm{W}(t))^{T}\nabla f(\bm{W}(t))$ denotes the Frobenius inner product of two matrices,
  $\nabla f(\bm{W}(t))=\left[\nabla f(\bm{w}^{(1)}(t)),\nabla f(\bm{w}^{(2)}(t)),\cdots,\nabla f(\bm{w}^{(M)}(t))\right]$ is the gradient of $f(\bm{W})$ at point $\bm{W}(t)$, and $l$ is a step size.
As a result of simple calculation, we get
\begin{multline}
\nabla f(\bm{w}^{(m)}(t))=(\bm{X}^{(m)})^{T}(\bm{X}^{(m)}\bm{w}^{(m)}(t)-\bm{y})\\
+2\sum_{k=1}^{M}(\bm{X}^{(m)})^{T}\bm{L}_{m,k}\bm{X}^{(k)}\bm{w}^{(k)}(t),
\end{multline}
where $\bm{L}_{m,k}\in\mathbb{R}^{N\times N}$ denotes the $(m,k)$-th block of $\bm{L}$ in (\ref{xxlap}), i.e., $\bm{L}_{m,k}=[L_{i,j}]_{\substack{1+(m-1)N\leq i\leq mN\\1+(k-1)N\leq j\leq kN}}$.

By ignoring the terms independent of $\bm{W}$ in (\ref{update1}), the update procedure can be written as
\begin{equation}\label{update2}
\bm{W}(t+1)=\argmin_{\bm{W}}\;\frac{1}{2}\lVert\bm{W}-\bm{V}(t)\rVert^{2}_{F}+lg(\bm{W}),
\end{equation}
where $\bm{V}(t)=\bm{W}(t)-l\nabla f(\bm{W}(t))$. In fact, (\ref{update2}) is equivalently expressed as
\begin{equation}
\bm{W}(t+1)=\text{\textbf{prox}}_{lg}(\bm{V}(t)),
\end{equation}
where $\text{\textbf{prox}}_{lg}$ denotes the proximal operator \cite{Parikh1} of the scaled function $lg$. Due to the separability of $\bm{W}(t+1)$ on each row, i.e., $\bm{W}^{i}(t+1)$, in (\ref{update2}), we can solve the optimization problem for each row individually:
\begin{equation}\label{update3}
\bm{W}^{i}(t+1)=\argmin_{\bm{W}^{i}}\;\frac{1}{2}\lVert\bm{W}^{i}-\bm{V}^{i}(t)\rVert_2^{2}+l\beta\lVert\bm{W}^{i}\rVert_2.
\end{equation}
In (\ref{update3}), the closed-form solution of $\bm{W}^{i}(t+1)$ can be easily obtained \cite{Parikh1}:
\begin{equation}
\bm{W}^{i\ast} =
  \begin{cases}
  \left(1-\frac{l\beta}{\lVert\bm{V}^{i}(t)\rVert_2}\right)\bm{V}^{i}(t)      & \quad \text{if } \lVert\bm{V}^{i}(t)\rVert_2\geq l\beta\\
    \bm{0}  & \quad \text{otherwise}.
  \end{cases}
\end{equation}

Furthermore, in order to accelerate the proximal gradient method, we introduce an auxiliary variable
\begin{equation}\label{qvariable}
\bm{Q}(t)=\bm{W}(t)+\frac{\alpha(t-1)-1}{\alpha(t)}(\bm{W}(t)-\bm{W}(t-1))
\end{equation}
and compute the gradient descent based on $\bm{Q}(t)$ instead of $\bm{W}(t)$, where the coefficient $\alpha(t)$ is set as
\begin{equation}
\alpha(t)=\frac{1+\sqrt{1+4\alpha(t-1)^2}}{2}.
\end{equation}

The pseudocode of the proposed optimization algorithm is summarized in Algorithm \ref{algo}.

\begin{algorithm}[!t]
\caption{}

\vspace{2mm}

\textbf{Input:} the data $\{\bm{X}^{(m)}\}_{m=1}^{M}$ and the response vector $\bm{y}$;\\
\textbf{Output:} $\bm{W}$;
\begin{algorithmic}[1]\label{algo}
\STATE \textbf{Initialization:} $t=1$, $\alpha(0)=1$, $l_0=1$, $\sigma=0.5$, $\bm{W}(0)=\bm{W}(1)=\bm{0}$, $\beta$, $\gamma$, $\lambda$;
\STATE \textbf{for} $t=1$ to \textit{Max-Iteration}, \textbf{do}
\STATE \quad Computer $\bm{Q}(t)$ by (\ref{qvariable})
\STATE \quad $l=l_{t-1}$
\STATE \quad \textbf{while} $f(\bm{W}(t+1))+g(\bm{W}(t+1))>\Omega_l(\bm{W}(t+1)),\bm{Q}(t))$,\\
        \quad\quad\quad\quad here $\bm{W}(t+1)$ is computed by (\ref{update2}), \textbf{do}
  \STATE \quad\quad $l=\sigma l$
  \STATE \quad \textbf{end while}
  \STATE \quad $l_t=l$
  \STATE \textbf{end for}
  \STATE \textbf{if} convergence \textbf{then}
  \STATE \quad $\bm{W}=\bm{W}(t+1)$, terminate
  \STATE \textbf{end if}
\end{algorithmic}
\end{algorithm}

\section{Experimental Results}\label{sec3}

\subsection{Data preprocessing}
In this study, we used the Philadelphia Neurodevelopmental Cohort (PNC) dataset \cite{pncdata1,pncdata2} for performance evaluation. The PNC is a large-scale collaborative research project between the Brain Behavior Laboratory at the University of Pennsylvania and the Center for Applied Genomics at the Children¡¯s Hospital of Philadelphia. The primary objective of the PNC project was to characterize brain and behavior interaction with genetics that combines neuroimaging, diverse clinical and cognitive phenotypes, and genomics. Nearly $900$ adolescents aged $8$--$22$ years underwent multimodal neuroimaging including resting-state fMRI, and fMRI of working memory and emotion identification tasks (called nback fMRI and emotion fMRI, respectively) in this research. All data acquired as part of the PNC can be freely downloaded from the public dbGaP site (\url{www.ncbi.nlm.nih.gov/projects/gap/cgi-bin/study.cgi?study_id=phs000607.v1.p1}).

\begin{figure}[!h]
  \centering
\includegraphics[width=1\columnwidth]{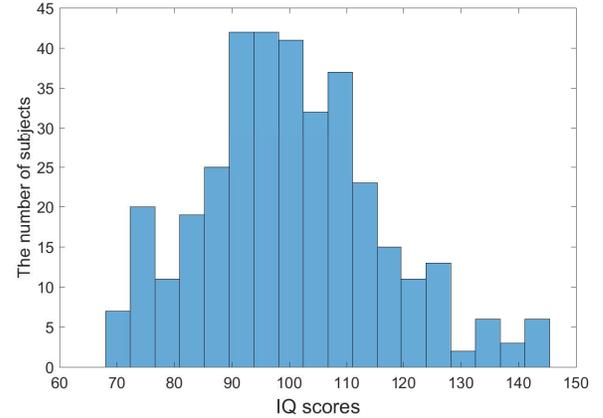}\\
  \caption{The IQ score distribution among the $355$ subjects.}\label{fig_IQ}
\end{figure}

We investigated the relationship between individual differences in IQ and brain activity during the engagement of cognitive abilities i.e., working memory and emotion identification, in this paper. The IQ scores of subjects were assessed with the Wide Range Achievement Test (WRAT), which was one test from a $1$-hour computerized neurocognitive battery (CNB) administered in the PNC. The WRAT is a standardized achievement test to measure an individual's learning ability, e.g., reading recognition, spelling, and math computation \cite{Wilkinson}, and hence provides a reliable estimate of IQ. To mitigate the influence of age over the final results, we excluded subjects whose ages were below $16$ years \cite{Zille1}. As a consequence, we were left with $355$ subjects (age: $16$--$22$ and $\text{mean}=18.21$ years; WRAT score: $70$--$145$ and $\text{mean}=100.57$; female/male: $204/151$), providing both nback fMRI and emotion fMRI. The distribution of IQ scores of these subjects is shown in Fig. \ref{fig_IQ}.

\begin{figure*}[!t]
  \centering
  \includegraphics[width=2\columnwidth]{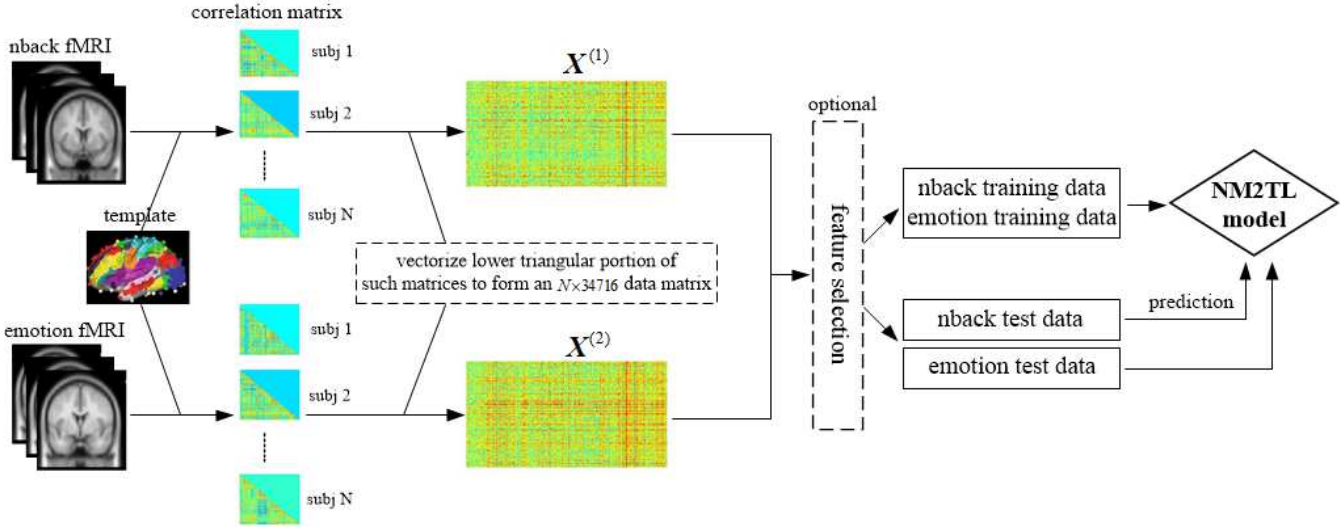}\\
  \caption{The flowchart of the proposed framework in this study.}\label{fig_flowchart}
\end{figure*}

All MRI scans were performed on a single $3$T Siemens TIM Trio whole-body scanner. In the fractal $n$-back task to probe working memory, subjects were required to respond to a presented fractal only when it was the same as the one presented on a previous trial. In the emotion identification task, subjects were asked to identify $60$ faces displaying neutral, happy, sad, angry, or fearful expressions.
All image data were acquired with a single-shot, interleaved multi-slice, gradient-echo, echo planar imaging sequence. We implemented image preprocessing separately for nback fMRI and emotion fMRI of the selected $355$ subjects. The preprocessing procedures were similar to those used in \cite{Zille1,Fang1,Hu1,lxiao2018}. Specifically, standard preprocessing steps were applied using SPM12 (\url{www.fil.ion.ucl.ac.uk/spm/}), which primarily consisted of motion correction, spatial normalization to standard MNI space, and spatial smoothing with a $3$mm FWHM Gaussian kernel. The functional time courses were subsequently band-pass filtered at $0.01$--$0.1$Hz. $264$ ROIs were defined to describe the whole brain as $10$mm diameter spheres centered upon ROI coordinates introduced in \cite{Power1}. We then calculated the Pearson correlation between the time courses of each pair of ROIs, resulting in a $264\times264$ correlation matrix (FC matrix) for each subject in each single fMRI modality (here we regarded our modalities as fMRI data collected under the two paradigms). To avoid repeated information, only the lower triangular portion of the symmetrical correlation matrix was properly reformed into a vector with $34716$ correlation values. Fisher's z-transform was applied to these correlations to ensure normality. The $34716$ FCs (Fisher's z-transformed values) were the features used in all subsequent analysis. As a result, we extracted $34716$ features from nback fMRI and $34716$ features from emotion fMRI for each subject.

\begin{table*}[!t]
\centering
\renewcommand\arraystretch{1.75}
\caption{The comparison of regression performance of nback fMRI and emotion fMRI by different predictive models.}
\label{t_result}
\begin{threeparttable}
\setlength{\tabcolsep}{3.5mm}{
\begin{tabular}{cccccc}
\hline
Model                 &       & \textit{CC} (mean $\pm$ std) & $p$-value & \textit{RMSE} (mean $\pm$ std) & $p$-value \\ \hline
\multirow{2}{*}{SM}   & nback &   $0.3181\pm0.0187$    & $<0.001$          &  $15.3882\pm0.1546$   & $<0.001$   \\ \cline{2-6}
                      & emotion &  $0.3240\pm0.0144$   & $0.0033$       &   $15.3060\pm0.1051$  &  $<0.001$  \\ \hline
\multirow{2}{*}{MTL}  & nback &      $0.3217\pm0.0183$     &   $0.0026$  &  $15.3063\pm0.1507$  &    $<0.001$ \\ \cline{2-6}
                      & emotion &  $0.3222\pm0.0175$    &  $0.0043$         & $15.2645\pm0.1669$   &    $<0.001$  \\ \hline
\multirow{2}{*}{M2TL}  & nback &   $0.3348\pm0.0118$    &  $0.0458$   &    $14.9881\pm0.1238$ & $0.0070$    \\ \cline{2-6}
                      & emotion &  $0.3308\pm0.0139$    &   $0.0337$    &  $15.0056\pm0.1635$  &   $0.0353$         \\ \hline
\multirow{2}{*}{NM2TL} & nback &   $0.3472\pm0.0141$   &   --      &    $14.8251\pm0.0714$  &     --    \\ \cline{2-6}
                      & emotion &   $0.3443\pm0.0125$    &   --      &  $14.8658\pm0.1047$        &    --       \\ \hline
\end{tabular}}
\begin{tablenotes}
\vspace{0.1cm}
        \item[1] $p$-values were calculated by pairwise t-test comparisons between the regression accuracy of our NM2TL model and other competing models for each modality.
        \item[2] std denotes the standard deviation.
    \end{tablenotes}
    \end{threeparttable}
\end{table*}

\begin{figure*}[!b]
  \centering
  %\hspace{-1.2cm}
\includegraphics[width=1.6\columnwidth]{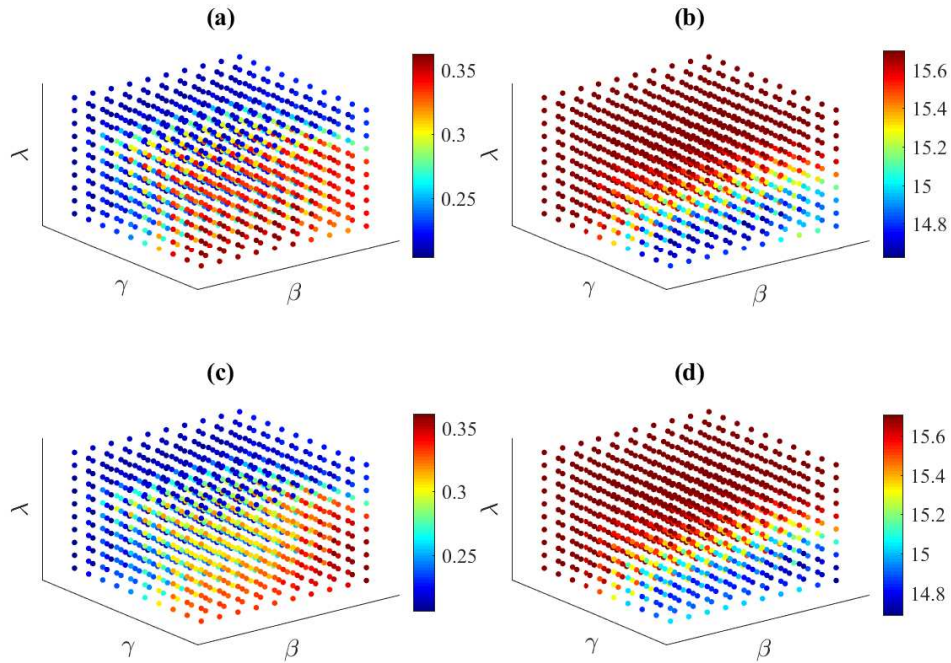}\\
  \caption{The regression performance of the proposed NM2TL model on different parameters' settings, i.e., $\beta,\gamma,\lambda\in\{10^{-3},3\times10^{-3},10^{-2},3\times10^{-2},10^{-1},0.3,1,3,10,30\}$. (a) nback fMRI (\textit{CC}s). (b) nback fMRI (\textit{RMSE}s). (c) emotion fMRI (\textit{CC}s). (d) emotion fMRI (\textit{RMSE}s).}\label{parameter}
\end{figure*}

\subsection{Experimental settings}
In our experiments, we compared the performance of the proposed NM2TL model and three other competing models: (1) SM (denoted as single modality based model with LASSO \cite{Tibshirani2011}, which is used to detect a significant subset of FCs from nback or emotion fMRI); (2) MTL \cite{Zhang1}; and (3) M2TL \cite{Jie1}.
We used a $5$-fold cross-validation (CV) technique to evaluate the IQ prediction performance of all these predictive models.
That is, the whole set of subjects was first randomly partitioned into $5$ disjoint subsets of as nearly equal size as possible; then each subset was successively selected as the test set and the other $4$ subsets were used for training the predictive model; and finally the trained model was applied to predict IQ scores of the subjects in the test set. This process was repeated for $10$ times independently to reduce the effect of sampling bias in the CV. All regularization parameters in the models, including the group-sparsity level $\beta$ and the manifold regularization parameters $\gamma$ and $\lambda$, were tuned by a $5$-fold inner CV on the training set through a grid search within their respective ranges, i.e., $\beta,\gamma,\lambda\in\{10^{-3},3\times10^{-3},10^{-2},3\times10^{-2},10^{-1},0.3,1,3,10,30\}$. The $K$ in the $K$-nearest neighbor rule for the graph similarity matrix calculation was empirically set as $10$.

One of the challenges encountered when using these predictive models is that whole-brain FC data consist of a large number of features (i.e., FCs) and a relatively small number of samples (i.e., subjects). This would give rise to various issues, such as proneness to overfitting, difficult interpretability, and computational burden. To this end, we used a simple univariate feature filtering technique to reduce the
number of features prior to inputting into the predictive models. Specifically, we discarded features for which the $p$-values of the correlation with IQ scores of subjects in the nback and emotion fMRI training set were both greater than or equal to $0.05$, and then trained the predictive models. All the remaining features of training subjects were normalized to have zero mean and unit norm, and the estimated mean and norm values of training subjects were used to normalize the corresponding features of testing subjects. Accordingly, we also conducted the mean-centering on IQ scores of training subjects and then used the mean IQ value of training subjects to normalize the IQ scores of testing subjects. The model performance on each modality was quantified as the root mean square error (\textit{RMSE}) and the correlation coefficient (\textit{CC}) between predicted and actual IQ scores of subjects in the test set. An overview of the proposed framework was outlined in Fig. \ref{fig_flowchart}.

\begin{figure*}[!t]
  \centering
\includegraphics[width=2\columnwidth]{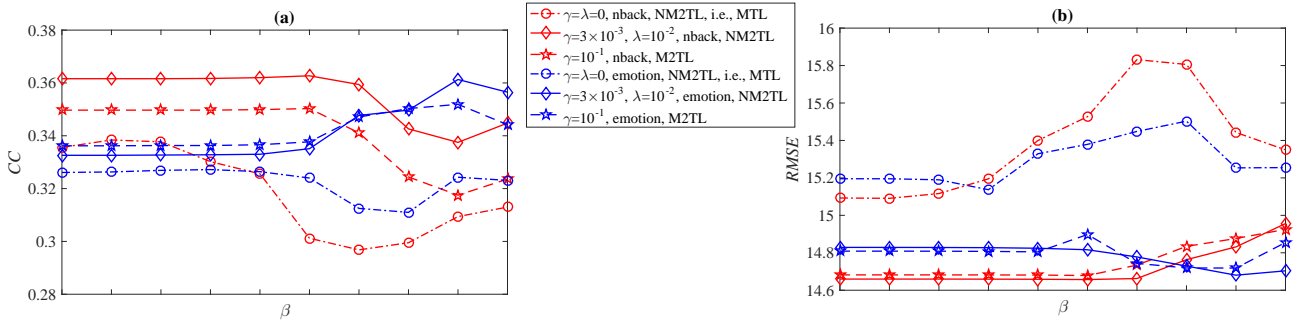}\\
  \caption{The regression performance with respect to the values of $\beta$, i.e., $\beta\in\{10^{-3},3\times10^{-3},10^{-2},3\times10^{-2},10^{-1},0.3,1,3,10,30\}$, and the selection of $\gamma$ and $\lambda$. (a) the performance of nback fMRI and emotion fMRI in terms of the \textit{CC}. (b) the performance of nback fMRI and emotion fMRI in terms of the \textit{RMSE}.}\label{compare}
\end{figure*}

\subsection{Regression results}

Table \ref{t_result} summarizes the regression performance of all competing models for IQ prediction.
As we can see from Table \ref{t_result}, the proposed NM2TL model consistently outperformed the other predictive models in terms of both the \textit{RMSE} and the \textit{CC}. Specifically, our proposed NM2TL model achieved the best \textit{CC}s of $0.3472$ for nback fMRI and $0.3443$ for emotion fMRI, and the best \textit{RMSE}s of $14.8251$ for nback fMRI and $14.8658$ for emotion fMRI. The next best performance was obtained by the M2TL model, i.e., $0.3348$ for nback fMRI and $0.3308$ for emotion fMRI in terms of the \textit{CC}, and $14.9881$ for nback fMRI and $15.0056$ for emotion fMRI in terms of the \textit{RMSE}. As shown in Table \ref{t_result}, the MTL model, which utilized the multi-task learning for a joint analysis of two modalities (tasks), achieved mostly better regression performance than the single-task based model (i.e., the SM model). It suggests that it is beneficial to use the multi-task learning for integrating complementary information from multiple modalities by jointly selecting a sparse set of common features. In addition, the manifold regularizers in the M2TL and the NM2TL models that can exploit the structure information of data still help increase the performance. Specifically, the proposed NM2TL model outperformed the MTL model, improving the performance by $0.0255$ and $0.0221$ in the \textit{CC}s, and by $0.4812$ and $0.3987$ in the \textit{RMSE}s, for nback fMRI and emotion fMRI, respectively. Meanwhile, in Table \ref{t_result} we reported the $p$-values of pairwise t-test based on the results of the $5$-fold CV to show statistically significant improvement of our proposed model.
In light of the fact that the best performance over the IQ regressions was all obtained by our proposed NM2TL model, we can well demonstrate that the designed manifold regularizer in our proposed model was effective in identifying more discriminative features associated with IQ. Therefore, it is shown that from the machine learning point of view, properly using different regularizers in the least square regression model has been proven as a valid way to circumvent the overfitting problem and find a compact solution,
especially in the high feature-dimension and low sample-size scenarios (e.g., in the field of neuroimaging analysis).

We next investigated the parameters' sensitivity by changing the values of $\beta,\gamma,\lambda$ in (\ref{NM3R model}). The results in Fig. \ref{parameter} show that the three parameters interactively affected the final performance, and our model was sensitive to them within only a small range.
For better understanding the effect of these parameters, we also presented the performance of the MTL model as baseline that does not include any manifold regularization term. It is worth noting that when $\gamma=\lambda=0$, our proposed NM2TL model will be degraded to the MTL model. As we can observe from Fig. \ref{compare}, our proposed NM2TL model and the M2TL model both consistently outperformed the MTL model (baseline) under all values of $\beta$. It can further embody the advantage of adding the manifold regularization term on top of the classical MTL model. Moreover, Fig. \ref{compare} shows that for each selected value of $\gamma$ and/or $\lambda$, the curve representing the performance with respect to different values of $\beta$ was very smooth as long as $\beta\leq10^{-1}$, which indicates that our proposed NM2TL model and the M2TL model were very robust to $\beta$ when $\beta$ lies in the range of small values.

\subsection{Discussion and future work}
Human intelligence can be broadly defined as the ability of comprehending and successfully responding to a wide
variety of factors in the external environment \cite{Neisser1996}. Also, IQ scores can be related to performance on cognitive tasks. Therefore, it is reasonable to examine the relationship between individual variations in IQ and brain activity during the engagement of the two cognitive tasks (i.e., working memory and emotion identification) in this paper. In the following, based on our proposed NM2TL model, we investigated the potential of both brain FCs and ROIs as biomarkers that are highly related to IQ, respectively.

\begin{figure}[!h]
 \centering
\includegraphics[width=1\columnwidth]{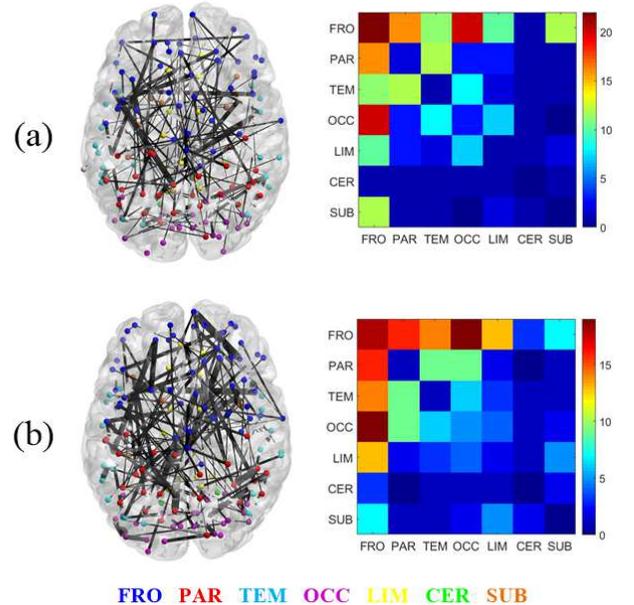}\\
  \caption{The visualization of the most discriminative $150$ FCs for (a) nback and (b) emotion modalities, respectively. The left are brain plots of functional graph in anatomical space, where the selected FCs are represented as the edges. The thicknesses of the edges consensus FCs with their weights. The ROIs are color-coded according to the cortical lobes: frontal (FRO), parietal (PAR), temporal (TEM), occipital (OCC), limbic (LIM), cerebellum (CER), and sub-lobar (SUB). The right are matrix plots that represent the total number of the selected edges connecting the ROIs across the cortical lobes.}\label{brain_FC}
\end{figure}

\begin{figure}[!h]
 \centering
\includegraphics[width=1\columnwidth]{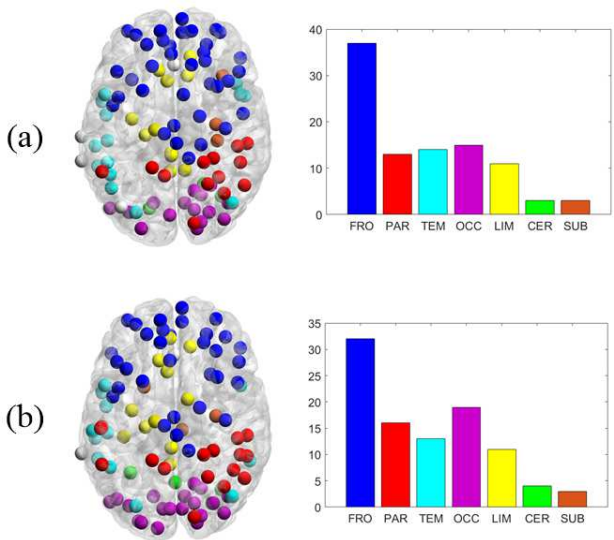}\\
  \caption{The visualization of the most discriminative $100$ ROIs for (a) nback and (b) emotion modalities, respectively. The left are brain plots of functional graph in anatomical space, where the selected ROIs are represented as the nodes. The ROIs are color-coded according to the cortical lobes. The right are bar plots that represent the total number of the selected ROIs in each cortical lobe.}\label{brain_node}
\end{figure}

To identify the most discriminative FCs, we averaged the obtained sparse regression coefficients $\bm{W}$ by these $5$-fold CV trials. The coefficient vector measures the relative importance of the FC features in predicting IQ scores. For ease of visualization, we selected $150$ nback FCs and emotion FCs with the largest averaged weights, respectively, and visualized them by using the BrainNet Viewer \cite{mingxia} in Fig. \ref{brain_FC}. It should be noted that these selected FCs were mainly located in frontal, parietal, temporal, and occipital lobes, which are in accordance with the previous studies in the literature. For instance, in \cite{tem1,tem2}, temporal lobe dysfunction has been shown to be related to attention-deficit/hyperactivity disorder (ADHD), which is significantly correlated with IQ impairments. Several regions within frontal, parietal, temporal, and occipital lobes have been identified as significant predictors of IQ in \cite{Greene2018,Gao2018,ooo1}. Also, to extract the most discriminative ROIs, we computed the ROI weights by summing the weights across all FCs for each ROI. In Fig. \ref{brain_node}, we visualized $100$ ROIs with the greatest relative prediction power on IQ for nback and emotion modalities, respectively. The results show that the largest number of the selected ROIs were located in frontal lobes, and the second largest number of the selected ROIs were in occipital, parietal, or temporal lobes. Interestingly, we also found that there were as many as $72$ overlapping ROIs between these two sets of $100$ ROIs selected separately from the corresponding two modalities.

In this paper, we focused on only two functional imaging modalities (here our
modalities refer to fMRI data collected under multiple paradigms), i.e., nback fMRI and emotion fMRI collected under two paradigms. The PNC dataset also includes resting-state fMRI. An interesting future work is to incorporate all three modalities (i.e., three types of fMRI data from different paradigms) together by means of the proposed NM2TL model, which may extract more discriminative information across modalities and further improve the IQ regression performance \cite{newadd2}. Another interesting note perhaps should be pointed out that the similarity measure of the data regardless of within each single modality or between modalities could largely affect the contribution of the manifold regularizer to the regression performance. Therefore, in order to reveal the intrinsic structure information inherent in multiple modalities, finding an effective and powerful strategy to learn the similarity of the data would be a high priority for improving our model.

\section{Conclusion}\label{sec4}
In this paper, based on the general linear regression model, we proposed a new manifold regularized multi-task learning model for a joint analysis of multiple modalities. Instead of including all high-dimensional features to predict performance, our proposed model was devised in such a way that the most prominent features that are able to influence performance with improved prediction accuracy can be successfully identified. In our proposed model, besides employing the group-sparsity regularizer to jointly select a small number of common features across multiple modalities (tasks), we designed a novel manifold regularizer to preserve the structure information both within and between modalities, which will most likely affect the final performance. Furthermore, we validated the effectiveness of our proposed model on the PNC dataset by using
fMRI based FC networks in two task conditions for IQ prediction.
The experimental results showed that our proposed model achieved superior performance in IQ prediction compared with other competing models. Moreover, we discovered IQ-relevant biomarkers in line with previous reports which may account for a proportion of the variance in human intelligence.

\end{document}